\documentclass[twocolumn,english,twocolumn,showpacs,preprintnumbers,amsmath,amssymb,superscriptaddress]{revtex4}
\usepackage[latin1]{inputenc}
\usepackage[symbol]{footmisc}
\DefineFNsymbols{nu}{* 1}\setfnsymbol{nu}
\usepackage{verbatim}
\usepackage{amsmath}
\usepackage{graphicx}
\usepackage{esint}

\makeatletter

\providecommand{\tabularnewline}{\\}

\@ifundefined{textcolor}{}
{%
 \definecolor{BLACK}{gray}{0}
 \definecolor{WHITE}{gray}{1}
 \definecolor{RED}{rgb}{1,0,0}
 \definecolor{GREEN}{rgb}{0,1,0}
 \definecolor{BLUE}{rgb}{0,0,1}
 \definecolor{CYAN}{cmyk}{1,0,0,0}
 \definecolor{MAGENTA}{cmyk}{0,1,0,0}
 \definecolor{YELLOW}{cmyk}{0,0,1,0}
}

\usepackage{babel}

\usepackage{babel}

\usepackage{babel}

\makeatother

\usepackage{babel}
\begin{document}

\title{Anisotropic emission of direct photons and thermal dileptons from Au+Au collisions
at $\sqrt{s_{NN}}=200$~GeV with EPOS3}

\author{Sheng-Xu Liu}

\affiliation{Key laboratory of Quark and Lepton Physics (Ministry of Education), 
Central China Normal University, Wuhan, China}

\author{Fu-Ming Liu \footnote{Correspongding author: liufm@mail.ccnu.edu.cn}}

\affiliation{Key laboratory of Quark and Lepton Physics (Ministry of Education),
Central China Normal University, Wuhan, China}

\author{Klaus Werner}

\affiliation{Laboratoire SUBATECH, University of Nantes - IN2P3/CNRS - Ecole desMines,
Nantes, France}

\author{Meng Yue}

\affiliation{Key laboratory of Quark and Lepton Physics (Ministry of Education), 
Central China Normal University, Wuhan, China}

\date{\today}
\begin{abstract}
Electromagnetic probes such as direct photons and dileptons are crucial to study the properties
of a Quark-Gluon Plasma (QGP) created in heavy ion collisions.  
Based on the (3+1)-dimensional event-by-event viscous hydrodynamic model EPOS 3.102, 
we calculated the anisotropic emission of thermal photons and dileptons in Au+Au collisions 
at the Relativistic Heavy Ion Collider (RHIC) energy $\sqrt{s_{NN}}=200$~GeV.
Thermal emissions from both QGP phase and hadronic gas phase are considered, with 
AMY rate for photons and Lattice QCD based rates for dileptons in QGP phase. 
For emission from hadron gas phase, rates based vector meson dominant model are used 
for both photons and dileptons. Non-thermal contribution to direct photons is calculated with 
next to leading order QCD. STAR cocktail data are directly used for non-thermal contribution to dileptons.
With the same space-time evolution of the collision systems, the two penetrating probes, photons and dileptonsshow some consistency, ie, the emission of both thermal photons and dileptons are underestimated,
compared with the transverse momentum spectra of direct photons measured by PHENIX collaboration and  
the invariant mass spectra of dileptons measured by STAR collaboration, for all centrality classes. 
With a good constraint of anisotropy of the plasma via the elliptic flow and triangular flow of charged hadrons, the resulted elliptic flow and triangular flow of direct photons agree with PHENIX measurements reasonably well. Thus we made predictions to the flows of thermal dileptons.  
The elliptic flow of thermal dileptons is predicted larger than the available results from other models, 
and comparable to the STAR measurement referring to all dileptons (thermal + cocktail) for minimal bias 
collisions. 
\end{abstract}
\maketitle

\section{INTRODUCTION}
Electro-magnetic probes such as direct photons and dileptons are considered as golden probes to a quark-gluon plasma (QGP)~\cite{{Shuryak}}.  Because of the small electro-magnetic coupling, they are expected to penetrate the     
hot dense matter created in relativistic heavy ion collisions and to carry us the information of the hot dense matter.  
The observed large elliptic flow of direct photons was once a puzzle in relativistic heavy ion physics~\cite{PHENIX2}. 
Our previous work provided a possibility to this puzzle with the delayed QGP formation in relativistic heavy ion collisions~\cite{FML2014}, 
reveals that direct photons, different from bulk hadrons, are able to carry the information of the system at the very early stage. 
Recently the McGill group made a nice reproduction of the elliptic flow, 
with a detailed investigation of the effects from viscosity~\cite{Paquet:2015lta}.
 
In this work, we will investigate the transverse momentum spectra, elliptic flow $v_2$ and triangular flow $v_3$ of direct photons 
from Au+Au collisions at the Relativistic Heavy Ion Collider (RHIC) energy $\sqrt{s_{NN}}=200$~GeV,
without any additional parameter to EPOS3, a hydrodynamic model  which 
reasonably reproduces hadronic data such as rapidity distributions, 
transverse momenta, elliptic flows and triangular flows 
in this collision systems. 
The comparison to the measured direct photon data will be performed thereafter. 

Then based on the same description of the space-time evolution of the collision systems, we will investigate the anisotropic emission of 
thermal dileptons.  In additional to the thermal sources, dileptons may also be produced from non-thermal sources such as primordial 
Drell-Yan annihilation and electromagnetic final-state decays of long-lived hadrons. To account for these, STAR cocktail results 
from Refs~\cite{STAR_PRL} have been employed in this paper.  
The calculated invariant mass spectrum will be compared to the data measured by the STAR Collaboration~\cite{STAR_PRL}.
Then we will check the elliptic flow $v_2$ and high order coefficient $v_n$ of the thermal dielectrons from different   

The paper is organized as following. In section 2, we will introduce our calculation 
approach of direct photons and thermal dileptons, 
such as the space-time evolution of the collision system based on hydrodynamic model EPOS3, 
and thermal dilepton emission rates from QGP phase and hadronic phase. 
In section 3 we will present the results, first 
the relevant hadronic results such as elliptic flow $v_2$ and triangular flow $v_3$, 
then systematically the results of photons and dileptons.  
Finally, discussion and conclusion in section 4.

\section{CALCULATION APPROACH}
Both thermal photons and thermal dileptons are calculated as the integral of 
the emission rates over the space-time evolution of the collision systems.
So we first introduce the space-time evolution of the collision systems.
Then we introduce the integral formula and the rates.

\subsection{The space-time evolution of the collision system with EPOS3}
As explained in \cite{epos3}, EPOS3 is an event generator based on
a 3+1D viscous hydrodynamic evolution starting from flux tube initial
conditions, which are generated in the Gribov-Regge multiple scattering
framework \cite{eposbas}. An individual scattering is referred to
as Pomeron, identified with a parton ladder, eventually showing up
as flux tubes (or strings). Each parton ladder is composed of a pQCD
hard process, plus initial and final state linear parton emission.
Nonlinear effects are considered by using saturation scales $Q_{s}$,
depending on the energy and the number of participants connected to
the Pomeron in question.

The final state partonic system (corresponding to a Pomeron) amounts
to (usually two) color flux tubes, being mainly longitudinal, with
transversely moving pieces carrying the $p_{t}$ of the partons from
hard scatterings \cite{eposbas,epos2}. One has two flux tubes, based
on the cylindrical topology of the Pomerons, but each quark-antiquark
pair in the parton ladder will cut a string into two, in this sense
one may have more than two flux tubes. In any case, these flux tubes
constitute eventually both bulk matter, also referred to as ``core''
\cite{eposcore} (which thermalizes, flows, and finally hadronizes)
and jets (also referred to as ``corona''), according to some criteria
based on the energy of the string segments and the local string density.

Concerning the core, we use a 3+1D viscous hydrodynamic approach,
employing a realistic equation of state, compatible with lQCD results.
We employ for all calculations in this paper a value of $\eta/S=0.08$.
Whenever a temperature of $T_{H}=168~$MeV is reached, the
usual Cooper-Frye procedure is used to convert the fluid into particles. From
this point on, a hadronic cascade \cite{urqmd}, based on
hadronic cross sections is employed. 

So we have two phases. In QGP phase, the hydrodynamical expansion can naturally provide us
the complete space-time information, i.e. the four fluid velocity $u^{\mu}$ and temperature $T$ 
at any given space-time $x$, starting from some initial proper time $\tau_{0}$.
In the hadronic phase, we have from EPOS a complete description of hadron trajectories. We use
this to compute energy densities, flow velocities and net baryon density.
Assuming that the system may be approximated by a resonance gas in
equilibrium, we use the corresponding equation of state, c.f. Apendix
C of \cite{epos2}, to get the space-time information, such as the four fluid velocity $u^{\mu}$ and temperature $T$ at any given space-time $x$. 

In the following are shown some basic results of the space-time evolution for AuAu collisions 
at RHIC energy $\sqrt{s_{NN}}=200$~GeV.
\begin{figure}
\includegraphics[scale=0.8]{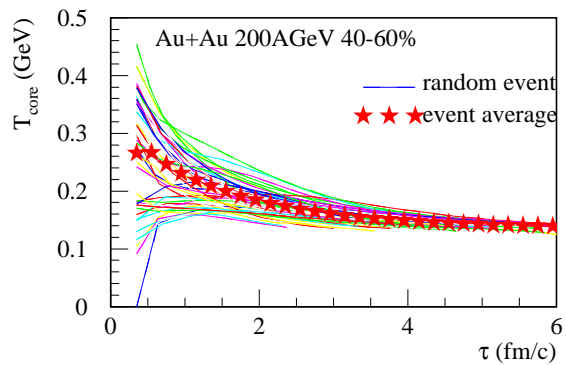}

 \caption{\label{Tcore} (Color Online) The time evolution of the temperature at the center point (x,y,z)=(0,0,0)
 in AuAu collisions at  $\sqrt{s_{NN}}=200$~GeV  with centrality 40-60\%.  Thin lines stand for random events. 
Stars for event averaged results.}
\end{figure}

First is shown the time evolution of the temperature at the center point (x,y,z)=(0,0,0) of the created hot dense matter 
in AuAu collisions at  $\sqrt{s_{NN}}=200$~GeV  with centrality 40-60\% in  Fig.~\ref{Tcore}.
Here the thin lines are results of some random EPOS3 events, stars for the averaged result over many events. 
In most events, the center temperature decreases with time. So does the event-averaged result. 
However, the core temperature fluctuates from event to event quite a lot.
In some events, the core temperature may increase during some short time range, being heated by the domain nearby.

  \begin{figure}
\includegraphics[scale=0.8]{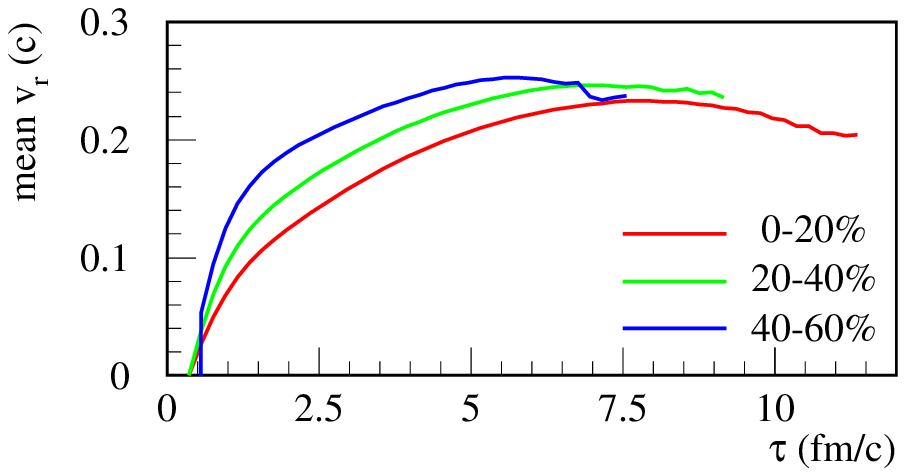}

 \caption{\label{vrm} (Color Online)  The time evolution of mean radial flow velocity in AuAu collisions at  $\sqrt{s_{NN}}=200$~GeV   
for three centrality classes, 0-20\%, 20-40\% and 40-60\%.    }
\end{figure}

Second in Fig.~\ref{vrm} is shown the time evolution of the mean radial velocity in the Au+Au systems 
at $\sqrt{s_{NN}}=200$~GeV for three centrality classes, 0-20\%, 20-40\% and 40-60\%. 
Radial flow $ v_r =\sqrt{v_x^2 +v_y^2} $ has some effect to the transverse momentum spectrum of produced particles. Moreover, 
it generates elliptic flow $v_2$, with its strength and asymmetry along azimuthal direction. 
Here is shown only the averaged strength at each given time $\tau$, the energy density weighted result over the whole space, then event average.
We can see the flow velocity is initially zero. Then it increases rapidly at the beginning then slowly saturates. 
The system of more central collisions has a longer life, but the upper limit of the mean radial velocity depends very weakly on centrality.

 \begin{figure}
\includegraphics[scale=0.8]{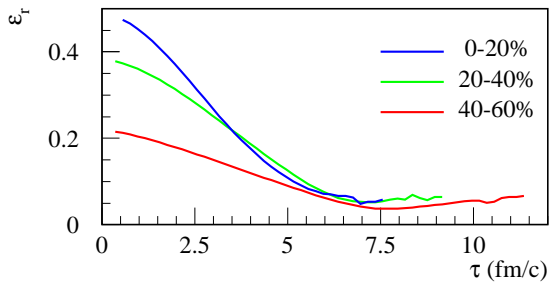}
 \caption{\label{ec2} (Color Online) The time evolution of the space eccentricity in AuAu collisions at  $\sqrt{s_{NN}}=200$~GeV
for three centrality classes, 0-20\%, 20-40\% and 40-60\%.   }
\end{figure}
Finally in Fig.~\ref{ec2} is shown the time evolution of the space eccentricity
 in AuAu collisions at  $\sqrt{s_{NN}}=200$~GeV
for three centrality classes, 0-20\%, 20-40\% and 40-60\%.
The space eccentricity is calculated as 
 \begin{equation}
\epsilon_r =  \frac { < y^2- x^2 >} {  < y^2 + x^2  >},
\end{equation}
 Here $ < ... > $
 stands for an energy density weighted space integral, additionally event average is also made. 
We can see the collision system initially carries a large eccentricity on average, because of the overlapping shape of the colliding nuclei. With the hydrodynamic expansion, the space eccentricity decreases. 

\subsection{Transverse momentum of direct photons and Invariant mass spectrum of dileptons}

The transverse momentum spectra of thermal photons can be written as 
\begin{equation}
\frac{dN}{dy\, d^{2}p_{t}}=\int d^{4}x\,\Gamma(E^{*},T)\label{eq1}
\end{equation}
where $\Gamma(E^{*},T)$ is the Lorentz invariant thermal photons
emission rate which covers the contributions from the QGP phase \cite{AMY}
and HG phase \cite{MYM} as our previous work, $d^{4}x=\tau\, d\tau\, dx\, dy\, d\eta$
being the volume-element, and the photon energy in the local rest frame
\begin{equation}
E^{*}=p^{\mu}u_{\mu}= \gamma p_{0}-\gamma\vec{p}\cdot\vec{v} \label{eq:lorentztr0}
\end{equation}
 with $p^{\mu}=(E, \vec{v})$ is the photon's four momentum in the laboratory frame, 
$u^{\mu}=(\gamma, \vec{v})$ is the four flow velocity, 
$v^{2}=\vec{v}\cdot\vec{v}$ and $\gamma=1/\sqrt{1-v^{2}/c^{2}}$. 
Temperature $T$ and  the fluid velocity $u^{\mu}=(\gamma,\vec v)$ is provided by EPOS3 as explained above.
The space-time integral starts from the default initial time of EPOS3, $\tau_0=0.35$~fm/c, 
until the end of the system evolution shown on average in Fig.~\ref{vrm}.  

The invariant mass spectrum of thermal dileptons reads
\begin{equation}
\frac{dN}{dM}(M)=\int d^{4}x\,\frac{Md^{3}q}{q_{0}}\,\Gamma'(T,q^*) \label{eq:intR}
\end{equation}
where $M$ and $q$ are the invariant mass and four momentum of the
dilepton pair in the laboratory frame, satisfying $M^{2}=q^{2}$. Again, 
the fluid velocity $u^{\mu}=(\gamma,\vec v)$ connects the four-momentum of the dilepton 
$q$ and $q^*$, in the laboratory frame and the local rest frame of the hot dense matter, respectively,via
\begin{equation}
q_{0}^{*}=\gamma q_{0}-\gamma\vec{q}\cdot\vec{v} \label{eq:lorentztr1}
\end{equation}
and
\begin{equation}
\vec{q^{*}}=\vec{q}+(\gamma-1)(\vec{q}\cdot\vec{v})\cdot\vec{v}/v^{2}-\gamma q^{0}\vec{v}.\label{eq:lorentztr2}
\end{equation}
The Lorentz invariant dilepton emission rates 
\begin{equation}
\Gamma'(T,q^*)=\frac{dR}{d^{4}q}(T,q^*)  
   =\frac{dN}{d^{4}xd^{4}q}(T,q^*) 
\end{equation}
will be explained in detail in the following, since this is our first dilepton paper.
The momentum integral ranges from 0 to 5GeV/c (results do not depend on the upper bound as long
as it is much bigger than the temperature).
 The STAR data are obtained with cuts on the single-electron tracks in the lab,
like $p_{t}>0.2$~GeV/c and $\left|\eta\right|< 0.9$.
These have a significant impact on both yields and shape of the low-mass dilepton spectrum. 
We simulate virtual photon with a given four-momentum decay into electron-positron pair, then obtain the acceptance efficiency of a virtual photon
according to the $\eta$  and $p_t$ distribution of produced electrons. Then put this acceptance factor into the space-time integral.

\begin{figure}
\includegraphics[scale=0.8]{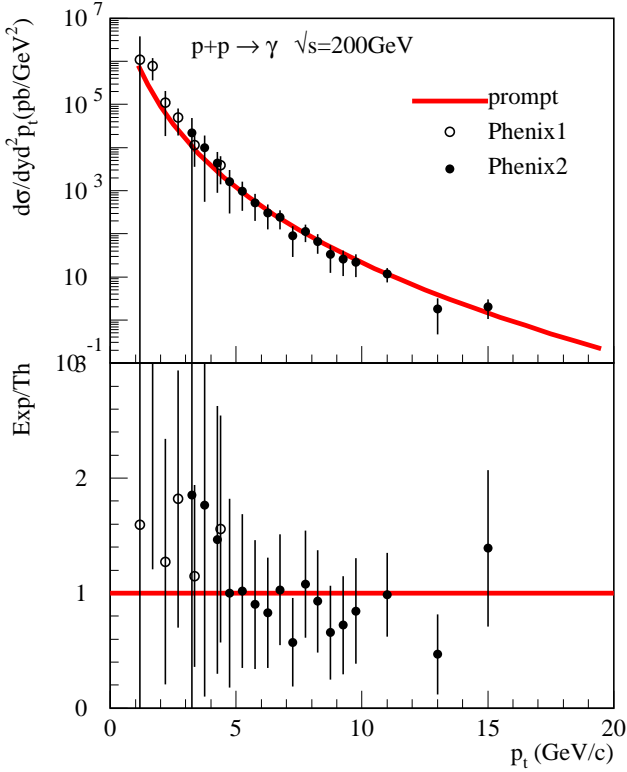}
 \caption{\label{pp200} (Color Online) Photon production from pp collisions at 
 $\sqrt{s}=200$~GeV. The solid line is our next to leading order QCD 
calculation for prompt photons. Dots are PHENIX data, full~\cite{PHENIXpp1} 
and empty~ \cite{Adare:2008ab}.}
\end{figure}

There are also non-thermal sources for direct photons and dileptons.
Non-thermal sources to dileptons such as primordial Drell-Yan annihilation and electromagnetic final-state decays of long-lived hadrons
are not calculated. Instead, STAR cocktail results from Refs~\cite{STAR_PRL} have been employed.
The non-thermal direct photon source is simplified as prompt photons, 
as a next to leading order QCD calculation in pp collisions~\cite{Liu:2011dk}.
Here we present the $p_t$ spectrum of prompt photons (red solid line) 
from pp collisions 
at $\sqrt s= 200$~GeV compared with PHENIX data~\cite{PHENIXpp1,Adare:2008ab}
 (dots) in ~\ref{pp200}.
A good coincidence in the hard region has been obained.
In heavy ion collisions, nuclear parton distribution functions (nPDF) $G_{a/A}(x)$ is used. 
For a nucleus with mass $A$, neutron number $N$, and proton number $Z$,
\begin{equation}
G_{a/A}(x)=\left[\frac{N}{A}G_{a/N}(x)+\frac{Z}{A}G_{a/P}(x)\right]R_{a}(x,A).\label{eq:PdisA}\end{equation}
where $G_{a/p}(x)$ is PDF in proton, $R_{a}$ factor is taken into account of shadowing effect and EMC effect~\cite{EKS98}.

\subsection{Dilepton Emission Rates }

This is our first work on dilepton. So it is necessary to explain 
thermal dilepton emission rates in QGP phase and HG phase in detail.
Photon emission rates have been employed in our previous work and kept.
For dileptons emitted from the QGP phase, asymptotic freedom implies
that the production rate in the intermediate invariant mass region
($1.<M<3.0GeV/c^{2}$) at high temperatures and densities can be described
by perturbation theory as in Ref.\cite{Rapp_1}. Recent progress using
thermal lattice QCD (lQCD) to calculate dilepton rates nonperturbatively
at vanishing three-momentum $q=0$ has been reported in Ref. \cite{Ding}.
For practical applications, an extrapolation to finite $q$ values
is needed. A construction proposed in Ref. \cite{Rapp_3} and employed
in our calculation reads 
\begin{equation}
\begin{split}\frac{d^{4}R}{d^{4}q}=\frac{\alpha^{2}}{4\pi^{4}}f^{B}(q_{0},T)C_{{\rm EM}}\{1+\frac{2T}{q}\ln[\frac{1+x_{+}}{1+x_{-}}]\\
+2\pi\alpha_{s}\frac{T^{2}}{M^{2}}K\,F(M^{2})\ln(1+\frac{2.912q_{0}}{4\pi\alpha_{s}T})\frac{2q_{0}^{2}+M^{2}}{3q_{0}^{2}}\},\label{eq:rateQGP}
\end{split}
\end{equation}
where $\alpha$ is the electromagnetic coupling constant, $f^{B}(q_{0},T)$
the thermal Bose distribution, $C_{{\rm EM}}\equiv\sum_{q=u,d,s}e_{q}^{2}$,
$x_{\pm}=\exp[-(q_{0}\pm q)/2T]$. The quantity $\alpha_{s}$ is the
temperature-dependent strong coupling constant, $K$ a constant factor
(equal 2), and finally we have a form factor $F(M^{2})=\frac{\Lambda^{2}}{\Lambda^{2}+M^{2}}$
with $\Lambda=2T$.

Dilepton emission Rate from hadronic gas is considered with the vector meson dominance model \cite{Gounaris}, where the
hadronic electromagnetic current operator is equal to the linear combination of the known neutral vector meson field operators, 
most notably V=$\rho$, $\omega$, $\phi$. This describes dilepton production successfully
\cite{Rapp_1}, where the dilepton emission rate via the vector meson V is \cite{Vujanovic} 
\begin{equation}
\frac{d^{4}R_{{\rm V}}}{d^{4}q}=-\frac{\alpha^{2}m_{{\rm V}}^{4}}{\pi^{3}M^{2}g_{{\rm V}}^{2}}{\rm Im}D_{{\rm V}}\frac{1}{e^{\frac{q_{0}}{T}}-1},\label{eq:rateHG}
\end{equation}
with the coupling constant $g_{{\rm V}}$ determined by the measured
decay rate of vector-meson to dilepton production in vacuum, and $m_{{\rm V}}$
being the mass of the vector-meson. The imaginary part of the retarded
vector meson propagator is given as 
\begin{equation}
{\rm Im}D_{{\rm V}}=\frac{{\rm Im}\Pi_{{\rm V}}}{(M^{2}-m_{{\rm V}}^{2}-{\rm Re}\Pi_{{\rm V}})^{2}+({\rm Im}\Pi_{{\rm V}})^{2}},\label{eq:ImD}
\end{equation}
where $\Pi_{{\rm V}}$ is the self-energy of the vector meson V.

We consider V=$\rho$ and neglect the complexity of $\omega$ and
$\phi$, which seems consistent with the STAR dilepton data taking\cite{STAR_PRL}.
The $\rho$ meson self-energy 
\begin{equation}
\Pi_{\rho}=\Pi_{\rho}^{vac}+\sum_{a}\Pi_{\rho a}\:,\label{eq:selfE2terms}
\end{equation}
for the contribution from the vacuum and from $\rho$ meson scattering
with hadrons of type $a$ in the hadronic gas, respectively.

The vacuum part $\Pi_{\rho}^{vac}$ is obtained from Gounaris-Sakurai
formula as Refs \cite{Gounaris,Kapusta}, which could describle the
pion electromagnetic form factor well, as measured in $e^{+}e^{-}$
annihilation Refs \cite{Strauch}: 
\begin{equation}
\begin{split}{\rm Re}\Pi_{\rho}^{vac}=\frac{g_{\rho}^{2}M^{2}}{48\pi^{2}}[(1-\frac{4m_{\pi}^{2}}{M^{2}})^{\frac{3}{2}}\ln\left|\frac{1+\sqrt{1-\frac{4m_{\pi}^{2}}{M^{2}}}}{1-\sqrt{1-\frac{4m_{\pi}^{2}}{M^{2}}}}\right|\\
+8m_{\pi}^{2}(\frac{1}{M^{2}}-\frac{1}{m_{\rho}^{2}})-2(\frac{p_{0}}{\omega_{0}})^{3}\ln(\frac{\omega_{0}+p_{0}}{m_{\pi}})],\label{eq:ReSelfE-vac}
\end{split}
\end{equation}

\begin{equation}
{\rm Im}\Pi_{\rho}^{vac}=-\frac{g_{\rho}^{2}M^{2}}{48\pi}(1-\frac{4m_{\pi}^{2}}{M^{2}})^{\frac{3}{2}}.\label{eq:ImSelfE-vac}
\end{equation}
Here, $2\omega_{0}=m_{\rho}=2\sqrt{m_{\pi}^{2}+p_{0}^{2}}$. The vacuum
width is $\Gamma_{\rho}^{vac}=\frac{g_{\rho}^{2}}{48\pi}m_{\rho}(\frac{p_{0}}{\omega_{0}})^{3}$.

The interactive part $\Pi_{\rho a}(E,p)$ is obtained from $\rho$
scattering from hadron of type $a$ in the hadronic gas 
\begin{equation}
\Pi_{\rho a}(E,p)=-4\pi\int\frac{d^{3}k}{(2\pi)^{3}}n_{a}(\omega)\frac{\sqrt{s}}{\omega}f_{\rho a}^{c.m.}(s),\label{eq:SelfE-inter}
\end{equation}
where $f^{c.m.}$ is the forward scattering amplitude in the c.m.
system, $E$ and $p$ are the energy and momentum of the $\rho$ meson,
$\omega^{2}=m_{a}^{2}+k^{2}$. The most copious hadrons in the hadronic
gas such as $\pi$ mesons ($a=\pi$) and nucleons ($a=N$) are considered
in our calculation. The quantity $n_{a}$ is the Bose-Einstein occupation
number of $\pi$ mesons and the Fermi-Dirac occupation number for
nucleons, with temperature $T$ and baryon chemical potential $\mu_{B}$
of the thermal bath provided by EPOS3 mentioned above. According to
Eq.(\ref{eq:lorentztr1}-\ref{eq:lorentztr2}), the emitted dileptons
will get a Lorentz boost with the collective flow velocity offered
by EPOS3, so the interactive term of self-energy and the resulted
emission rate are calculated in the rest frame of the thermal bath.

The $\rho a$ forward scattering amplitude in the center-of-mass frame
could be written as 
\begin{equation}
f_{\rho a}^{c.m.}=(1-x)f_{{\rm Res}}+xf_{{\rm Reg}}+f_{{\rm Pom}},
\end{equation}
where the Pomeron term is dual to the background upon which the resonances
are superimposed, $x$ is a function that matches the low energy Breit-Wigner
resonances and high energy Reggons (dual to s-channel resonances)
smoothly: 
\begin{equation}
x=0.5(1+\tanh(\frac{E_{\rho}-m_{\rho}-E_{\Delta}}{0.3})),
\end{equation}
where $E_{\Delta}$ is 1GeV and 4GeV for $\rho\pi$ and $\rho N$
scattering, respectively\cite{Xu1}.

Regge and Pomeron term have the same form \cite{Eletsky}: 
\begin{equation}
f_{{\rm Reg/Pom}}=-\frac{q_{c.m.}}{4\pi s}\frac{1+e^{-i\pi\alpha}}{\sin\pi\alpha}s^{\alpha}r^{\rho a},
\end{equation}
where the intercept $\alpha$ and residue $r^{\rho N}$, $r^{\rho\pi}$
are 0.642, 28.59, 12.74 for Regges and 1.093, 11.88, 7.508 for Pomerons,
respectively (the units yield a cross section in mb with energy in
GeV) .

The resonance term is \cite{Eletsky}: 
\begin{equation}
f_{{\rm Res}}=\frac{1}{2q_{c.m.}}\sum_{R}W_{\rho a}^{R}\frac{\Gamma_{R\to\rho a}}{M_{R}-\sqrt{s}-\frac{1}{2}i\Gamma_{R}},
\end{equation}
which involves a sum over a series of Breit-Wigner resonances of mass
$M_{R}$ and total width $\Gamma_{R}$. The resonances R used in our
calculation are listed in table I for $a=\pi$ and table II for $a=N$,
with R's name, mass, decay width, branching ratio, spin, isospin and
relative angular momentum Ref. \cite{Eletsky,Xu1}.

\begin{table}[!hbp]
\protect\caption{Meson resonances $R$ for $\rho\pi$ processes.}

\begin{tabular}{|c|c|c|c|c|c|c|}
\hline 
Name(R)  & Mass($M_{R}$)  & $\Gamma$  & BR  & S  & IS  & L \tabularnewline
\hline 
$\phi$(1020)  & 1.020  & 0.0045  & 0.13  & 1  & 0  & 1 \tabularnewline
\hline 
$h_{1}$(1170)  & 1.170  & 0.36  & 1  & 1  & 0  & 0 \tabularnewline
\hline 
$a_{1}$(1260)  & 1.230  & 0.40  & 0.68  & 1  & 1  & 0 \tabularnewline
\hline 
$\pi$(1300)  & 1.300  & 0.40  & 0.32  & 0  & 1  & 1 \tabularnewline
\hline 
$a_{2}$(1320)  & 1.318  & 0.107  & 0.70  & 2  & 1  & 2 \tabularnewline
\hline 
$\omega$(1420)  & 1.419  & 0.174  & 1  & 1  & 0  & 1 \tabularnewline
\hline 
\end{tabular}
\end{table}

\begin{table}[!hbp]
\protect\caption{Baryon resonances $R$ for $\rho N$ processes.}

\begin{tabular}{|c|c|c|c|c|c|c|}
\hline 
Name(R)  & Mass($M_{R}$)  & $\Gamma$  & BR  & S  & IS  & L \tabularnewline
\hline 
N(2090)  & 1.928  & 0.414  & 0.49  & 0.5  & 0.5  & 0 \tabularnewline
\hline 
N(1700)  & 1.737  & 0.249  & 0.13  & 1.5  & 0.5  & 0 \tabularnewline
\hline 
N(2080)  & 1.804  & 0.447  & 0.26  & 1.5  & 0.5  & 0 \tabularnewline
\hline 
N(2190)  & 2.127  & 0.547  & 0.29  & 3.5  & 0.5  & 2 \tabularnewline
\hline 
N(2100)  & 1.885  & 0.113  & 0.27  & 0.5  & 0.5  & 1 \tabularnewline
\hline 
N(1720)  & 1.717  & 0.383  & 0.87  & 1.5  & 0.5  & 1 \tabularnewline
\hline 
N(1900)  & 1.879  & 0.498  & 0.44  & 1.5  & 0.5  & 1 \tabularnewline
\hline 
N(2000)  & 1.903  & 0.494  & 0.60  & 2.5  & 0.5  & 1 \tabularnewline
\hline 
$\Delta$(1900)  & 1.920  & 0.263  & 0.38  & 0.5  & 1.5  & 0 \tabularnewline
\hline 
$\Delta$(1700)  & 1.762  & 0.599  & 0.08  & 1.5  & 1.5  & 0 \tabularnewline
\hline 
$\Delta$(1940)  & 2.057  & 0.460  & 0.35  & 1.5  & 1.5  & 0 \tabularnewline
\hline 
$\Delta$(2000)  & 1.752  & 0.251  & 0.22  & 2.5  & 1.5  & 1 \tabularnewline
\hline 
$\Delta$(1905)  & 1.881  & 0.327  & 0.86  & 2.5  & 1.5  & 1 \tabularnewline
\hline 
N(1520)  & 1.520  & 0.124  & 0.008  & 1.5  & 0.5  & 0 \tabularnewline
\hline 
$\Delta$(1232)  & 1.232  & 0.118  & 0.006  & 1.5  & 1.5  & 1 \tabularnewline
\hline 
\end{tabular}
\end{table}

\begin{figure}
\includegraphics[scale=0.4]{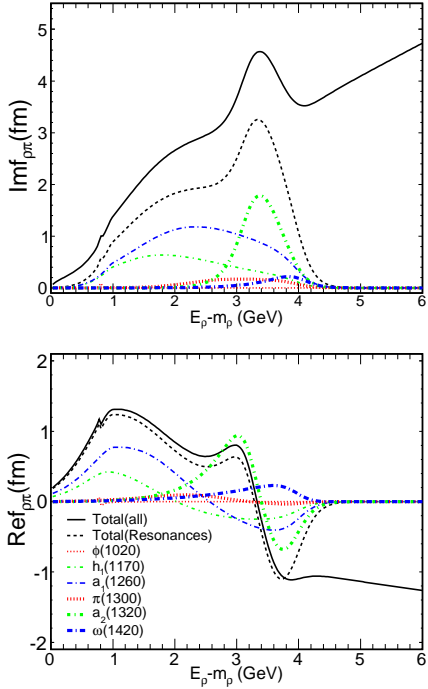}

\protect\caption{\label{fig:frhopi} (Color Online) The individual and total amplitude
for $\rho\pi\rightarrow R$ scattering (imaginary part: upper panel,
real part: lower panel).}
\end{figure}

\begin{figure}
\includegraphics[scale=0.4]{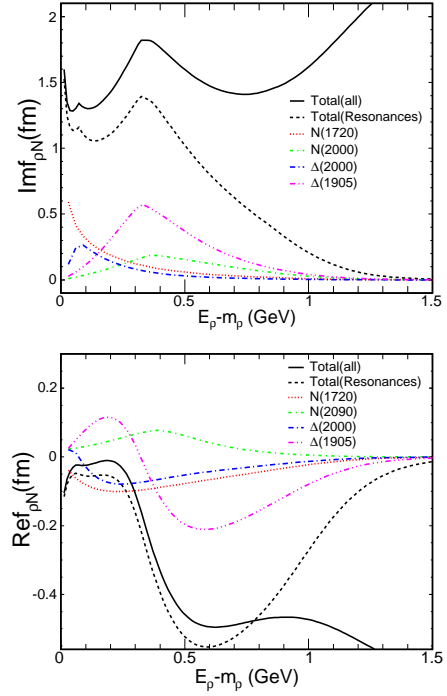}

\protect\caption{\label{fig:frhoN} (Color Online) The relevant individual and total
amplitude for $\rho N\rightarrow R$ scattering (imaginary part: upper
panel, real part: lower panel) .}
\end{figure}

The c.m. amplitude $f^{c.m.}$ and the scattering amplitude in the
rest frame of $a$, $f_{\rho a}$, are related by 
\begin{equation}
\sqrt{s}f_{\rho a}^{c.m.}(s)=m_{a}f_{\rho a}(E_{\rho}),\label{eq:frhoa}
\end{equation}
with $s=m_{\rho}^{2}+m_{a}^{2}+2E_{\rho}m_{a}$. The latter is plotted
in Fig.\ref{fig:frhopi} and Fig.\ref{fig:frhoN}, for $\rho\pi$
and $\rho N$ scattering, respectively. The total contribution (Resonances
+ Pomeron background + Regge) in both are shown as black solid lines,
and both are consistent with Ref.~\cite{Eletsky}. Each individual
process $\rho\pi\rightarrow R$ is also shown in Fig.\ref{fig:frhopi}.
Their sum is shown as black dashed line. We can see that the contribution
of resonances dominants the low energy scattering, and the dominant
channels are $R=a_{1}(1260)$ and $h_{1}(1170)$ due to their large
branch ratios. In Fig.\ref{fig:frhoN}, the relevant individual channels
$\rho N\rightarrow R$ are shown. The black dashed line is the total
baryon resonances contribution, among which $R=\Delta(1905),N(2000)$
and $N(1720)$ dominate both the imaginary and the real part of the
the amplitude for $\rho N$ scattering.

\begin{figure}
\includegraphics[scale=0.4]{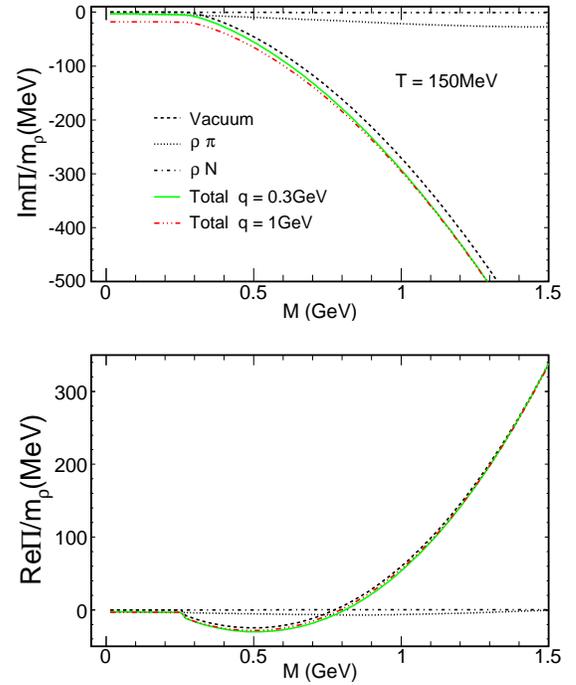}

\protect\caption{\label{fig:selfe} The self-energy of $\rho$-meson .}
\end{figure}

Putting the forward scattering amplitutes together, we get the interactive
term of self-energy of $\rho$-meson. The imgaginary and real parts
are shown in Fig.\ref{fig:selfe}, for a fixed $\rho$-meson
three-momentum $q=0.3$~GeV/c at $T=150$~MeV, with $\rho\pi$ scattering
dotted lines and $\rho N$ scattering dashed-dotted lines, respectively.
The contribution from vacuum is plotted as dashed lines. The total
one is plotted as solid lines. The vacuum contributes when
$M>2m_{\pi}$, then dominates over a wide region. 
To illustrate the $q$-dependence, the total contribution with $q=1$~GeV/c is also presented.  
Because of rho-pi scattering rather than rho-N scattering, the imaginary part of 
the self-energy  has an evident $q$-dependence at the low invariant mass.  

\begin{figure}
\includegraphics[scale=0.3]{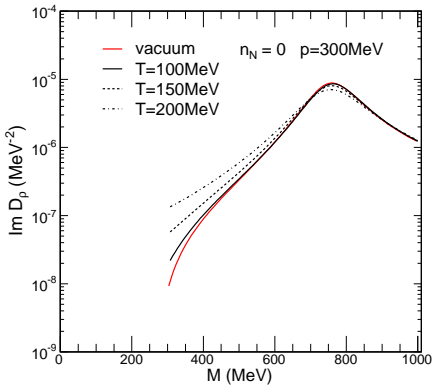}

\protect\caption{\label{fig:rhopropagator1} (Color Online) The imaginary part of the
$\rho$ meson propagators as a function of invariant mass for a momentum
of $300MeV/c$ and a nucleon density of $n_{N}=0$. Results are shown
for vacuum(red solid line) and three temperatures, $T=100MeV/c$ (black
solid line), $T=150MeV/c$ (black dashed line), $T=200MeV/c$ (black
dashed-dotted line).}
\end{figure}

Yet, the interaction with the medium gives a broadening effect to
the spectral density, $\rho(q)=-{\rm Im}D_{\rho}(q)/\pi$. In Fig.\ref{fig:rhopropagator1},
${\rm Im}D_{\rho}$ is plotted as a function of invariant mass $M$
for a $\rho$-meson momentum of $q=0.3$~GeV/c, in a thermal hadronic
gas of temperature T=100, 150 and 200MeV, respectively. The vacuum
contribution by itself is plotted as a red solid line. With the increase
of medium temperature, the $\rho$-meson scattering with hadron $a$
in the medium broadens the spectral density more and more, but the
peak remains close to $\rho$-meson mass. (Note: T=200MeV is only
to visualize the broadening effect, though it is too high for hadronic
gas. )

\begin{figure}
\includegraphics[scale=0.3]{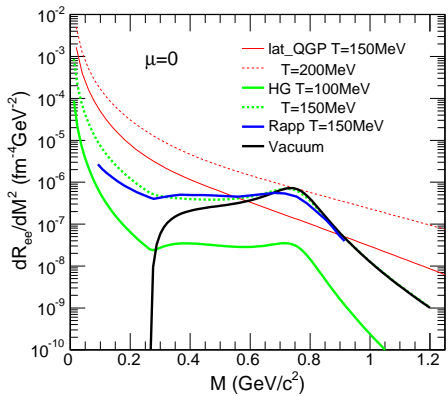}

\protect\caption{\label{fig:rate1} (Color Online) Emission rates of dielectrons from
QGP phase (red thin lines) at temperature: 150MeV (solid line), 200MeV
(dashed line), and HG phase (green thick lines) at temperature: 100MeV
(solid line), 150MeV (dashed line). The vacuum contribution to HG
rate at 150MeV is shown as black dashed line. HG rate from effective
interaction Lagrangians \cite{Rapp_4} at 150MeV is shown as blue
dashed-dotted line.}
\end{figure}

The emission rates at different phases are now obtained via Eq.(\ref{eq:rateQGP}-\ref{eq:rateHG}).
In Fig.\ref{fig:rate1}, the emission rate from the QGP phase is plotted
as red thin lines (dotted: 200MeV, solid: 150MeV), from the HG phase
as green thick lines (dotted: 150MeV, solid: 100MeV). A higher temperature
makes a stronger di-electron emission. Nevertheless the HG phase provides
a pronounced peak around the $\rho$-meson mass 775MeV which exceeds
the QGP contribution. This peak is essentially due to the vacuum term
in the self-energy (black dashed line). The emission rate from HG
phase at 150MeV obtained from effective interaction Lagrangians \cite{Rapp_4}
is also shown as a blue dotted dashed line. It is interesting to see
the two rates are quite close to each other, though different approaches
and channels are considered.

We should mention here that the above-mentioned emission rates work
typically for ideal hydrodynamics. For viscous hydrodynamics such
as EPOS3, a viscous correction is needed. The effect of shear viscosity
to the spectra and elliptic flow of dileptons in QGP phase and HG
phase has been investigated~\cite{Vujanovic:2012nq}. Similar work
has been done for direct photons~\cite{Shen:2013vja}. The elliptic
flow of dileptons seems more sensitive to viscous effect than their
spectra, however, it remains a modest effect.

\subsection{Elliptic flow $v_{2}$ and Triangular flow $v_{3}$}

The elliptic flow $v_{2}$ and the triangular one $v_{3}$ of thermal
photons/dileptons are calculated in a similar way as in Refs \cite{FML2014}.

For each event, the azimuthal angle dependence of the transverse momentum spectrum of thermal direct photons or 
of the invariant mass spectrum of thermal dielectrons can be decomposed into harmonics of the azimuthal
angle $\phi$ as 
\begin{equation}
\frac{dN}{d\phi} \sim\frac{1}{2\pi}[1+2v_{2}\cos2(\phi-\psi_{2})+2v_{3}\cos3(\phi-\psi_{3})+\cdots],\label{eq:dndphi}
\end{equation}
where $v_{2}$($v_{n}$) is the elliptic flow (higher order harmonics),
and $\psi_{n}$ is the $n$th-order event plane angle. Obviously,
$v_{n}$ and $\psi_{n}$ depend on the dielectron's invariant mass
$M$ and vary event by event. From Eq.~(\ref{eq:dndphi}), one can
easily get 
\begin{equation}
v_{e,n}\cos(n\psi_{e,n})=\frac{\int_{0}^{2\pi}\cos(n\phi)\frac{dN}{d\phi} |_e  d\phi}{\int_{0}^{2\pi}\frac{dN}{d\phi} |_e  d\phi}
\end{equation}
\begin{equation}
v_{e,n}\sin(n\psi_{e,n})=\frac{\int_{0}^{2\pi}\sin(n\phi)\frac{dN}{d\phi} |_e  d\phi}{\int_{0}^{2\pi}\frac{dN}{d\phi} |_e d\phi},
\end{equation}
with the subscript $e$ added to emphasis variables for a single event. 
Let's note their right sides as $<\cos n\phi>_e$ and $<\sin n\phi>_e$,
respectively. Then, for each event, the harmonics and reaction plane
of order $n$ can be obtained as 
\begin{equation}
v_{e,n}=\sqrt{ {<\cos(n\phi)>_e}^{2}+{<\sin(n\phi)>_e}^{2}}
\end{equation}
and 
\begin{equation}
\psi_{e,n}=\frac{1}{n}\arctan\frac{<\sin(n\phi)>_e}{<\cos(n\phi)>_e}.
\end{equation}
The elliptic flow $v_{2}$ and higher order harmonics $v_{n}$ of
the sample is obtained via event average
\begin{equation}
v_{n}= \sum_{e=1}^N \frac {v_{e,n}}{N}.      
\end{equation}
 This is equivalent to the definition in \cite{Gale:2012rq}. 

To obtain the flow coefficients $v_n$ of direct photons, the yield of prompt photons should be taken into the average.
Vanishing coefficients $v_n$ are assumed for prompt photons since collective motion has not yet involved at that time.
Because prompt photons dominant at high transverse momentum $p_t$ as seen in the following, 
the flow coefficients of direct photons vanish at high $p_t$.

\section{RESULTS}

\begin{figure}
\includegraphics[scale=0.8]{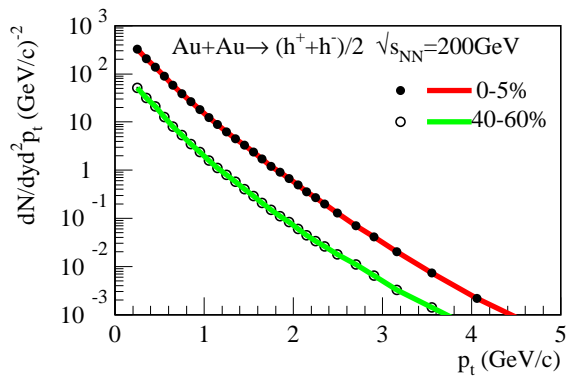}

 \caption{\label{pt} (Color Online) 
Transverse momentum spectra of charged hadrons from AuAu collisions at 
$\sqrt{s_{NN}}=200$~GeV  for centrality 0-5\% and  40-60\% simulated with EPOS3.102(lines).  Data points from STAR~\cite{Adams:2003kv}.   }
\end{figure}

\begin{figure*}
\includegraphics[scale=0.8]{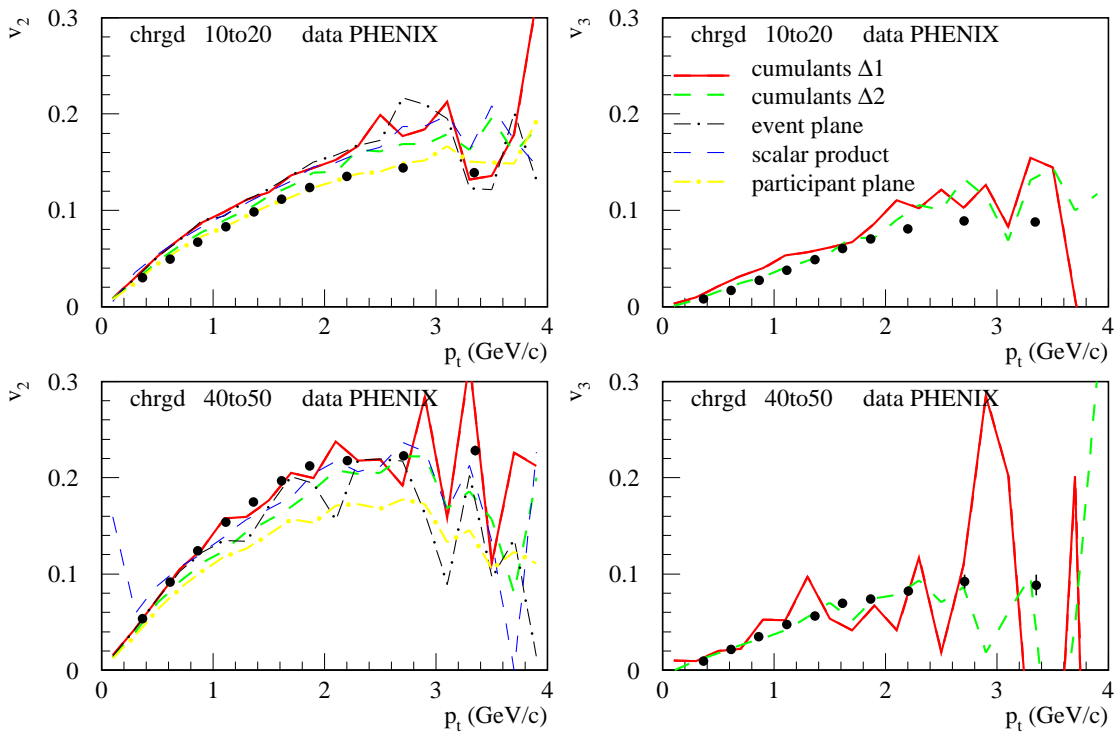}

 \caption{\label{gg2my06p1} (Color Online) Elliptic flow $v_2$ (left panels) and triangular flow $v_3$(right panels) of charged hadrons from AuAu collisions at $\sqrt{s_{NN}}=200$~GeV  for centrality 10-20\% (left) and  40-50\% (right).  Data points  from PHENIX~\cite{Adare:2011tg}.   }
\end{figure*}

First we check how well the hydrodynamic evolution is constrained by hadron data.  EPOS3 made a reasonable reproduction of the rapidity distributions and transverse momentum spectra of not only charged hadrons but also identified particles.
Here we show some relevant results. First the transverse mementum spectra of charged hadrons from AuAu collisions at 200GeV, with different centrality, as shown in Fig.~\ref{pt}. The lines are from EPOS3.102 simulation which coincide well 
with STAR data~\cite{Adams:2003kv}.
 In Fig.~\ref{gg2my06p1} are shown the elliptic flow $v_2$ (left panels) and triangular flow $v_3$ (right panels) of charged hadron from AuAu collisions at 200 GeV with centrality 10-20\% (upper panel) and 40-50\%  (lower panel).   
 The different types of curves stand for different approaches to obtain hadrons flow, ie,  cumulant approach with two particle pseudo-rapidity difference $\Delta \eta>1$ (red solid lines) , cumulant approach with $\Delta \eta > 2$  (green dashed lines) , event plane approach (black thin dashed dotted lines), scalar product approach (blue thin dashed lines), participant plane approach (yellow dashed dotted lines). 
The coincidence to experimental data\cite{Adare:2011tg} (full dots) reveals a reasonable constraint to the anisotropy of the collision systems.  

\begin{figure*}
\includegraphics[scale=0.8]{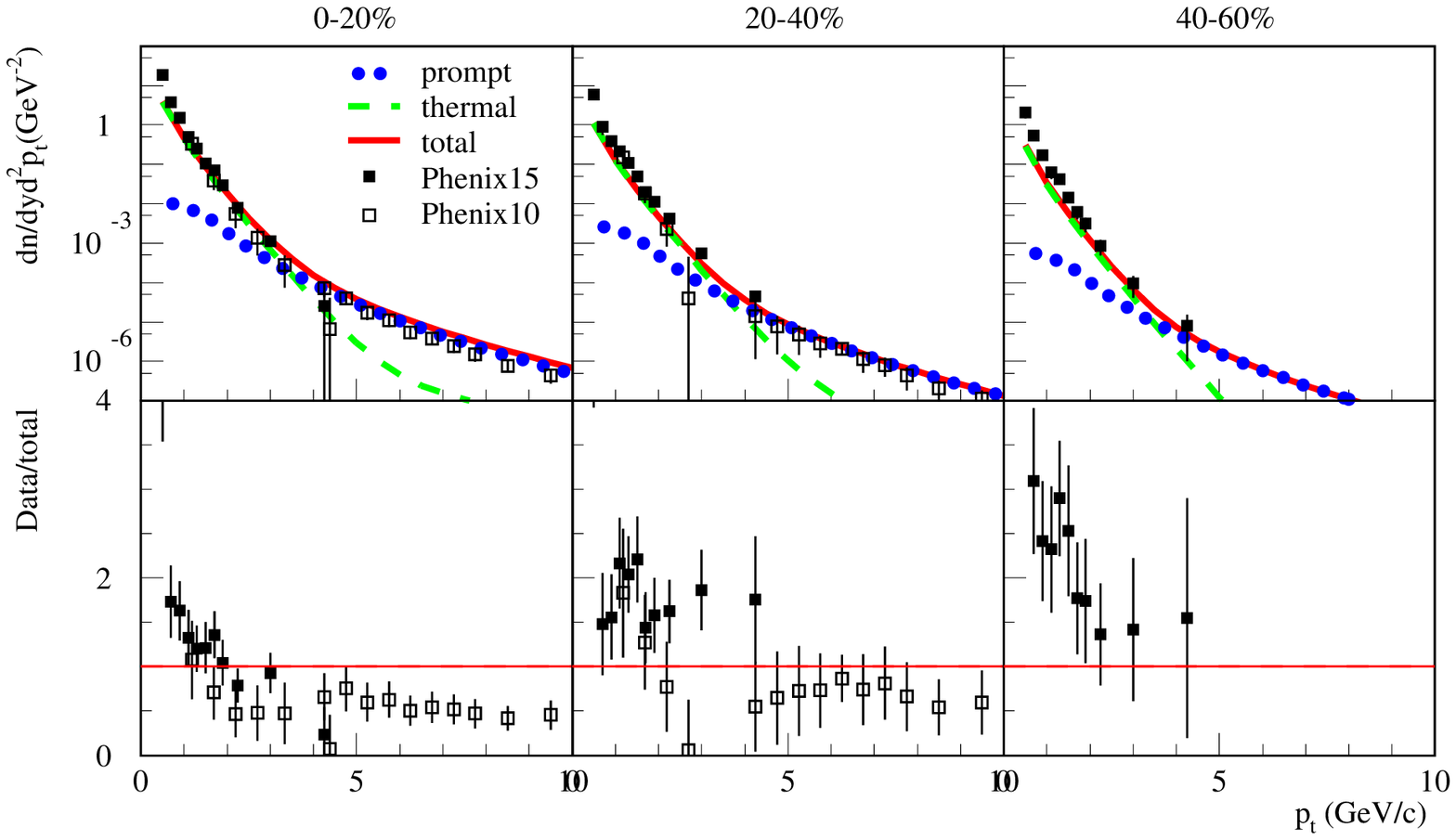}

 \caption{\label{200spectra} (Color Online) Upper panels: The transverse momentum spectra of photons
from AuAu collisions at $\sqrt{s_{NN}}=200$~GeV for centrality 0-20\%, 20-40\% and 40-60\% 
(from left to right). Total direct photons (solid lines) are decomposed into two sources, 
 prompt photons (dotted dashed lines) and thermal photons (dashed lines). 
Data points of direct photons from PHENIX~\cite{Adare:2008ab,Adare:2014fwh}. 
Lower panels: The ratio of PHENIX data to the calculated direct photons.}
\end{figure*}

Then we show the photon results, from the spectra to flows.
In Fig.~\ref{200spectra}, on the upper panels, the calculated transverse momentum spectra of direct photons (full solid lines) from AuAu collisions at three centrality classes, 
0-20\% (left panel), 20-40\%(middle panel) and 40-60\%(right panel), are decomposed to the two main sources, prompt (dotted lines) and thermal (dashed lines).
The dominance of the two sources to the transverse momentum regions is clear shown. 
The total contribution are compared to the PHENIX data~\cite{Adare:2008ab, Adare:2014fwh}.
This comparison is better shown in the lower panels. 
At high transverse momentum region, the prompt contribution is lower than data points constantly, 
for the centrality 0-20\%. So the scaled binary collision number in this centrality is better 
doubled in this centrality.  This shortcoming doesn't effect too much our judgement of the thermal emission, since the dominant $p_t$ regions of the two sources are quite clear.
At low transverse momentum region, the ratio Data/total is above unity, which means the thermal emission from our calculation is too low. The collision systems provided by EPOS3 do not shine enough,  
similar to the McGill results~\cite{Paquet:2015lta}.

\begin{figure*}
\includegraphics[scale=0.8]{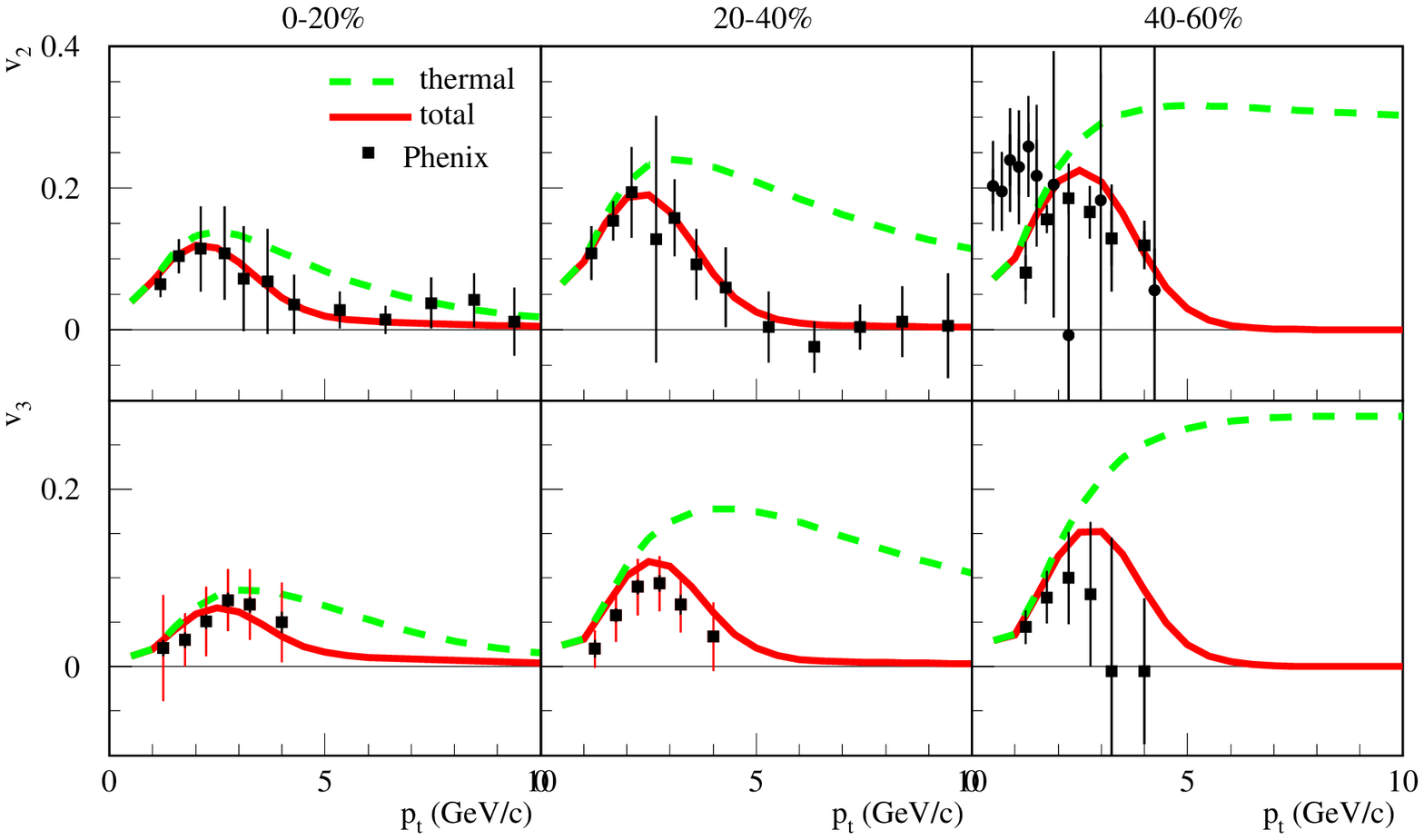}

 \caption{\label{v2v3} (Color Online) Elliptic flow $v_2$ (upper panels) and triangular flow $v_3$(lower panels) of direct photons (solid lines) and thermal photons (dashed lines) from AuAu collisions at $\sqrt{s_{NN}}=200$~GeV  for centrality 0-20\% , 20-40\% and 40-60\% (from left to right). Data points  $v_2$ and $v_3$ of direct photons from PHENIX~\cite{Bannier:2014bja}.   }
\end{figure*}

Now we present the anisotropic emission of photons.
In Fig.~\ref{v2v3}, the elliptic flow $v_2$ (upper panels) and triangular flow $v_3$ (lower panels) 
of photons (solid lines: direct photons) from AuAu collisions at 200GeV with centrality 0-20\%, 20-40\% and 40-60\% (from left to right)  are 
presented.  
The values of thermal photons are plotted as dashed lines. 
With the vanishing flows of prompt photons, the values of direct photons (red solid lines) are reduced, especially at high transverse momentum region. 
A good coincidence of the calculated elliptic flow to the PHENIX data\cite{PHENIX2}  (Full dots)  
is obtained for the three centrality classes. For triangular flow, the coincidence between experimental data and theoretical results is less good and centrality dependent. 

\begin{figure*}
\includegraphics[scale=0.8]{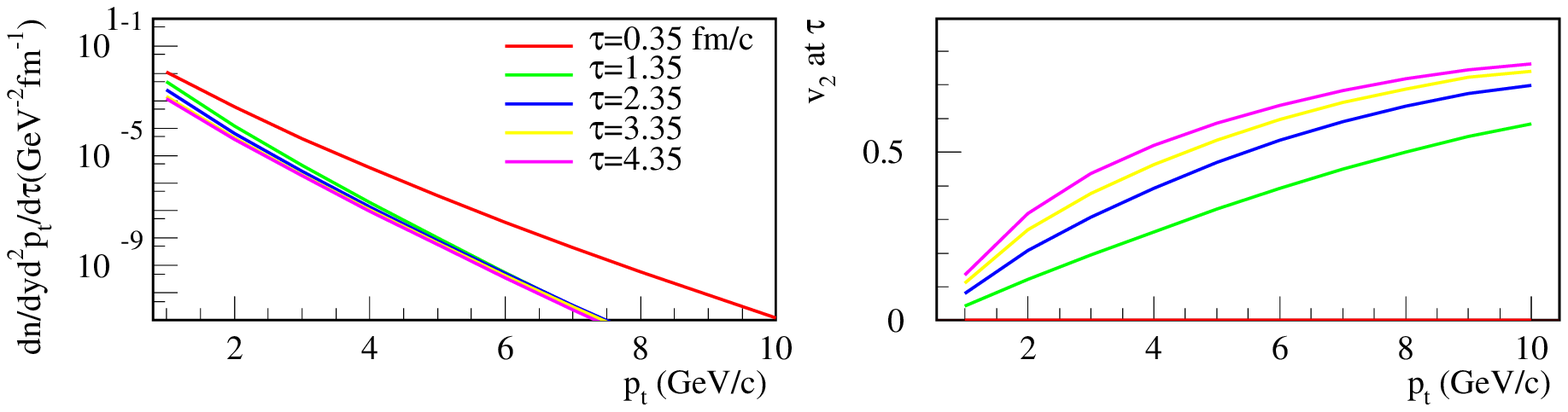}

 \caption{\label{cent60ta} (Color Online) 
The snapshots of the yields (left panel) and elliptic flow $v_2$ (right panel) of thermal photons
from AuAu collisions at $\sqrt{s_{NN}}=200$~GeV for centrality 40-60\%. Earlier emission makes more yield but smaller elliptic flow.}
\end{figure*}

An interesting question is to know how does the elliptic flow and triangular flow of photons are built up with time. 
Here in Fig.~\ref{cent60ta} some snapshots of the yields (left panel) and elliptic flow $v_2$ (right panel) of thermal photons are shown.
The increase of elliptic flow with time is evident. The elliptic flow at the initial time (red line) is zero
because of the vanishment of the initial radial flow velocity. 
However, the yields of the initial time (also red line) is the largest, despite the smallest system size.
High order coefficients $v_n$ behave similar to $v_2$. And so do the following thermal dileptons. 
The invariant mass effects not to the instantaneous flow coefficient, but to the instantaneous yields.
And the observables comparable to the experimental measurements are time-integrated results.  

\begin{figure*}
\includegraphics[scale=0.7]{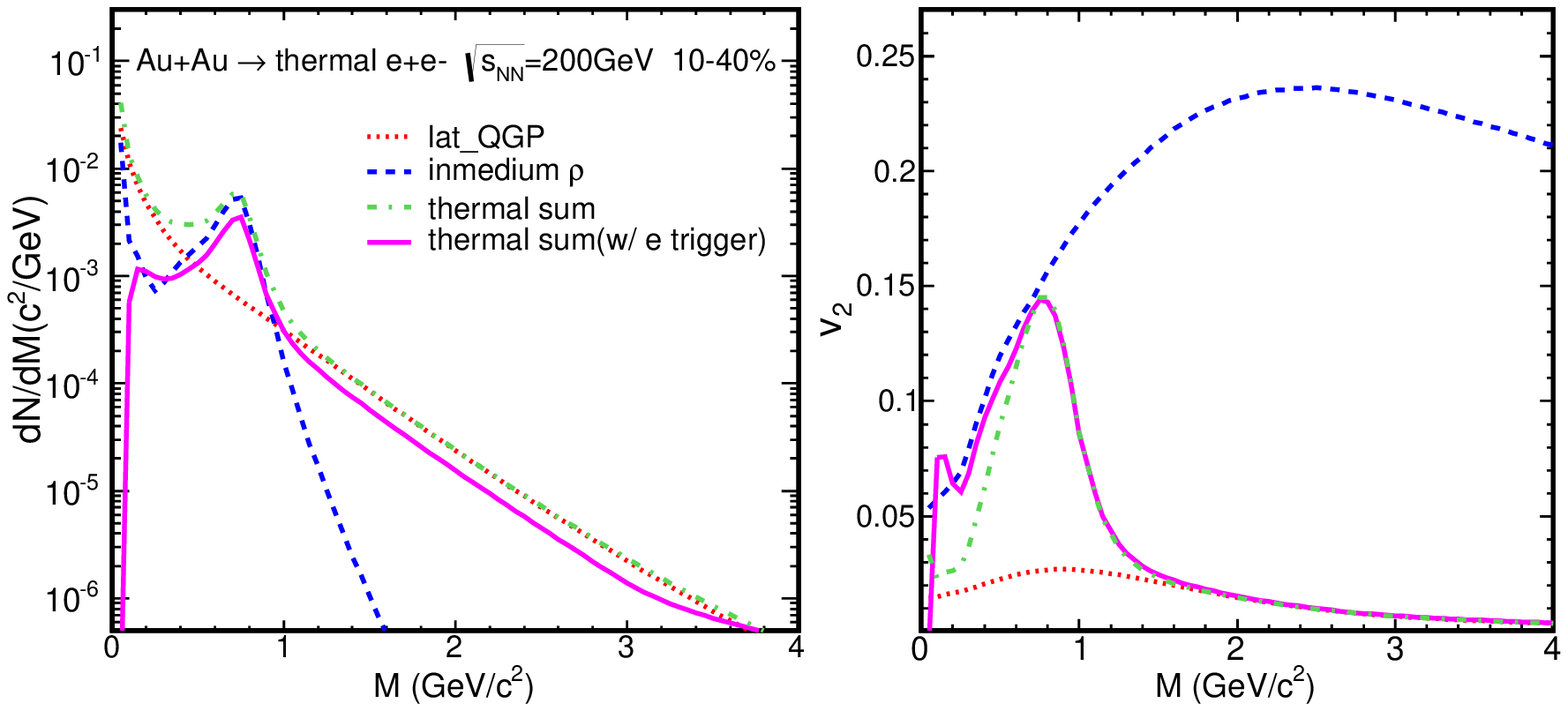}

\protect\caption{\label{fig:dndmandv2_10-40} (Color Online)
The invariant mass spectrum (left panel) and elliptic flow (right panel) of
 thermal di-electrons (green dotted dashed lines) from 
AuAu collisions at $\sqrt{s_{NN}}$=200~GeV 
with centrality 10-40\% are decomposed into di-electrons from QGP phase
(red dotted lines) and hadronic phase (blue dashed lines).
Solid lines are results with STAR single electron trigger. 
Following results without mention are all with this acceptance factor
due to single electron trigger.} 
\end{figure*}

Now we present the results of dileptons.
First, in Fig.~\ref{fig:dndmandv2_10-40}  
are shown the contributions from different phases, QGP phase (red dotted lines) and HG 
phase (blue dashed lines), to the invariant mass spectrum (left panel) and 
the elliptic flow (right panel) of the midrapidity thermal di-electrons from AuAu collisions at 
$\sqrt{s_{NN}}=200$~GeV with centrality 10-40\%. 
The QGP contribution dominates the thermal spectra at $M > 1$ ~GeV/c$^{2}$
region, as in the above figure on emission rates. And the peak from
the contribution of the hadronic phase indeed remains around the $\rho$
mass after the space-time integral of the emission rate. 
The elliptic flow of HG phase, the latter phase, is larger, consistent with our above snapshots. 
The contributions of the two phases together make the green dashed dotted lines.
Thus, in the total thermal spectrum, a pronounced peak still exists.
The elliptic flow ranges between the upper bound (values in hadronic phase) and 
lower bound (values in QGP phase), and is close to the corresponding bound of the dominating phase. 
To compare with STAR data, we have to take into account of the detector acceptance factor due 
to single electron trigger as mentioned above. 
This acceptance efficiency factor reduces the thermal contribution 
over a large region of invariant mass $M$, particularly strongest at the small $M$, c.f. the solid line on the left panel. Its effct to the elliptic flow is ignorable for large $M$, but very strong at small $M$, c.f. the solid line in the right panel. In the following, this acceptance factor is always included in our results. 

Next the invariant mass spectra is compared to STAR data. For the goal, non-thermal contribution such as STAR cocktail is included.
STAR cocktail contribution includes all the non-thermal contributions, such as contributions from Drell-Yan process and decay
of long lived hadrons, such as light mesons, D-mesons and so on. Based on the measurement of those hadrons, Tsallis fitting and decay simulation, provide us 
the invariant mass spectra of the cocktail contribution~\cite{STAR2}.

\begin{figure}
\includegraphics[scale=0.4]{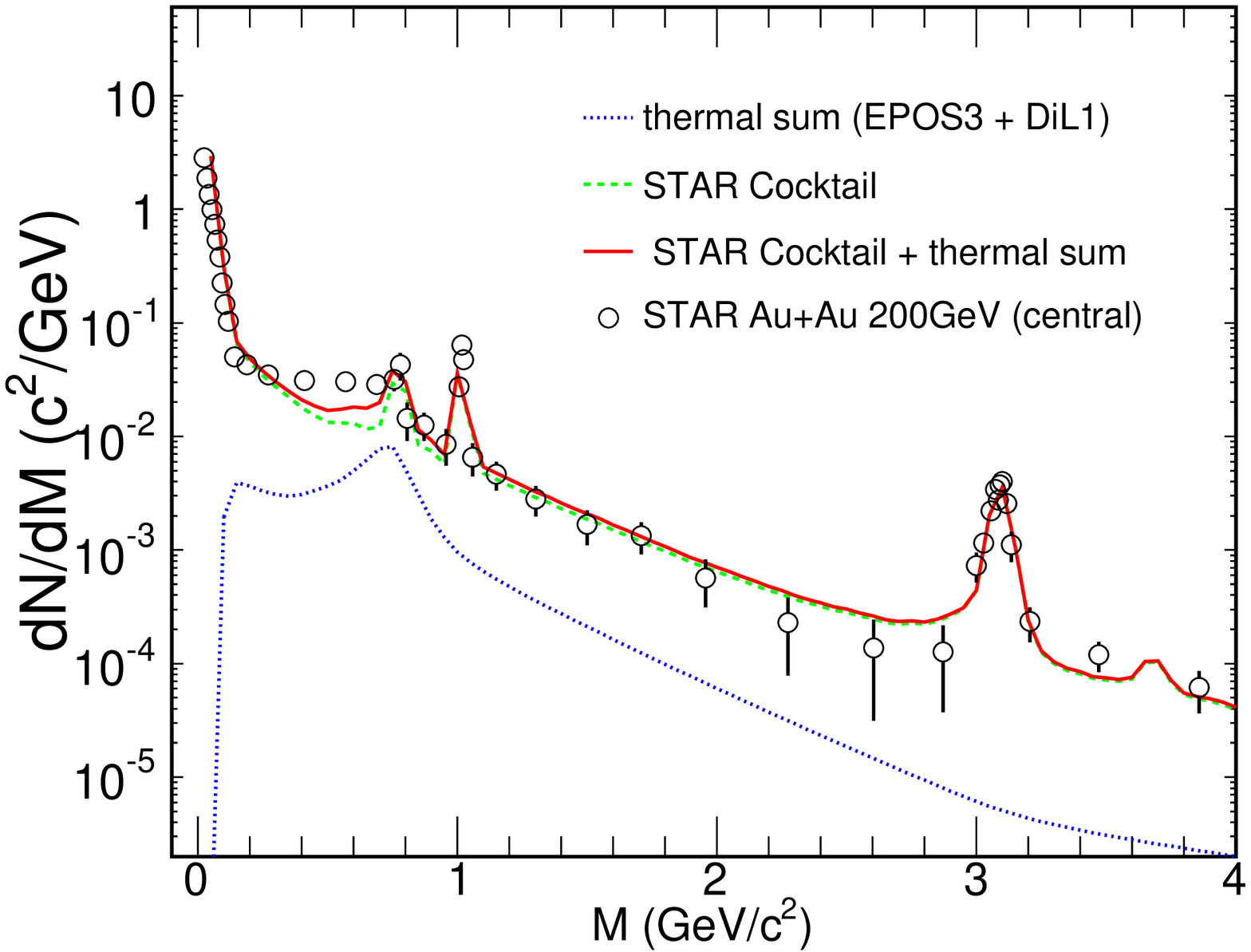}

\caption{\label{fig:dndm0-10} (Color Online) 
The invariant mass spectra of dielectrons from Au+Au collisions 
at $\sqrt{s_{NN}}=200$~GeV for centrality 0-10\% (solid line) is the sum of 
thermal contribution (dotted line) and STAR cocktail (dashed line). 
The measured invariant mass spectrum of dielectrons 
by STAR collaboration~\cite{STAR_PRL} are plotted as empty dots.}
\end{figure}

In Fig.\ref{fig:dndm0-10}, the invariant mass spectrum of dielectron 
from 0-10\% Au+Au central collisions at $\sqrt{s_{NN}}=200$~GeV 
( solid line ) is the sum of thermal contribution
(dotted line) and the STAR cocktail (dashed line).  
At intermediate and high invariant mass region, 
it coincides with the STAR measured result (empty dots), 
because of the common cocktail contribution.
At low invariant mass region, the calculated curve is lower than data points. 
More emission of thermal dilepton is needed to coincide with data, consistent to the photon results.  

\begin{figure*}
\includegraphics[scale=0.8]{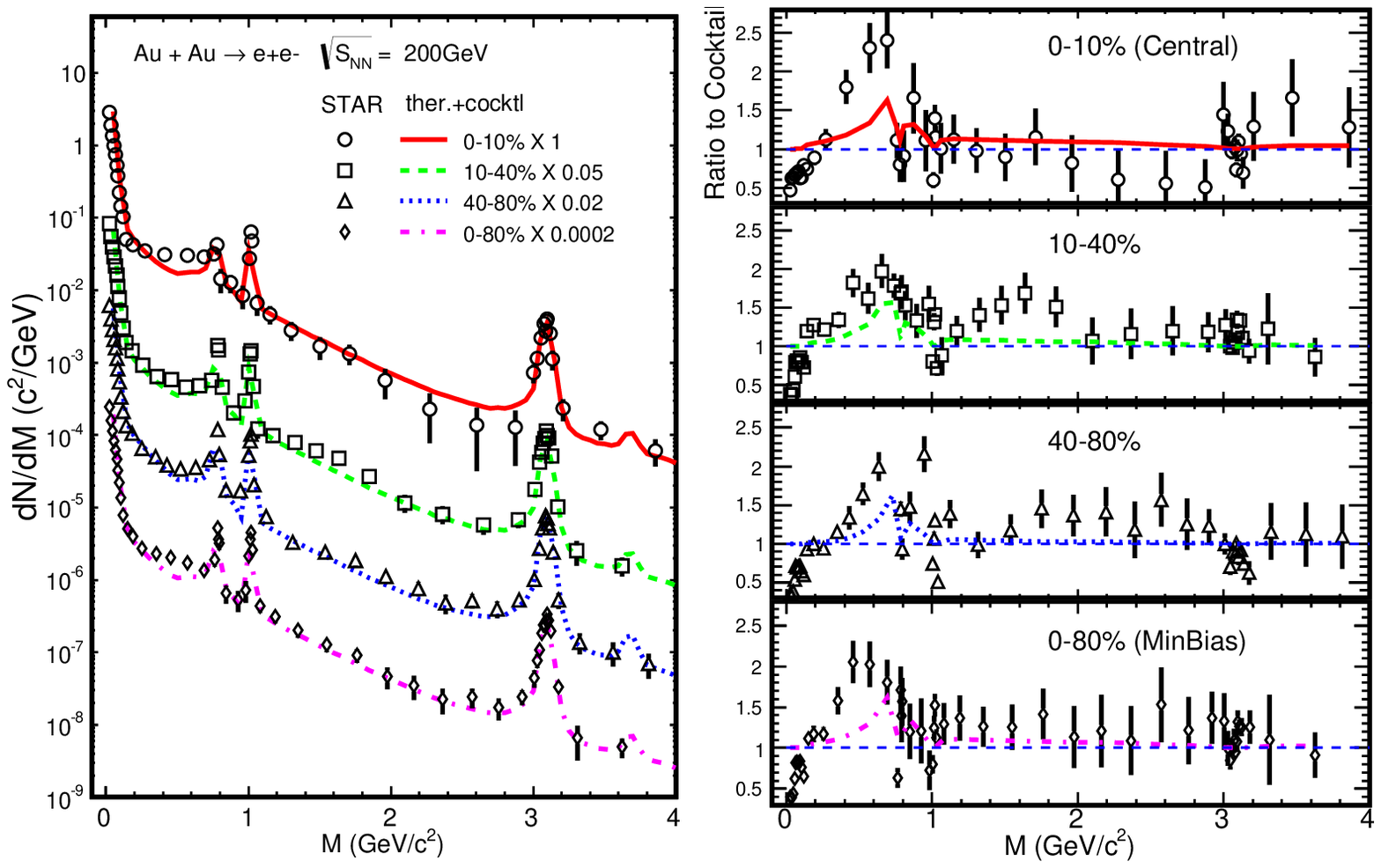}

\protect\caption{\label{fig:dndmandratio} (Color Online) 
Left panel: The invariant mass spectra of dielectrons (thermal+cocktail) 
are compared to STAR data \cite{STAR_PRL,STAR1},
for four centrality classes, 0-10\% (central), 10-40\%, 40-80\% and 
0-80\% (minimal bias) (from top to bottom). 
Right panel: The comparion is shown as the ratio to the cocktail contribution.} 
\end{figure*}

In Fig.\ref{fig:dndmandratio}(left panel) the invariant mass spectra
of dielectrons (thermal + cocktail) from Au+Au collisions at 
$\sqrt{s_{NN}}=200$~GeV for different centralities are compared to STAR data
\cite{STAR_PRL,STAR1}, from top to bottom: 0-10\% (central), 10-40\%,
40-80\% and 0-80\% (MinBias). For better visibility, the latter three
results are multiplied by factors: 0.05, 0.02 and 0.0002, respectively.
In the right panel, the comparison is better shown as the ratio to 
the cocktail contribution.
The calculated results are in fact lower than the data,
for all centralities, similar to the previous results of thermal photons.  
We miss about a half of the thermal dileptons to coincide with data, for all centralities. 
\begin{figure}
\includegraphics[scale=0.4]{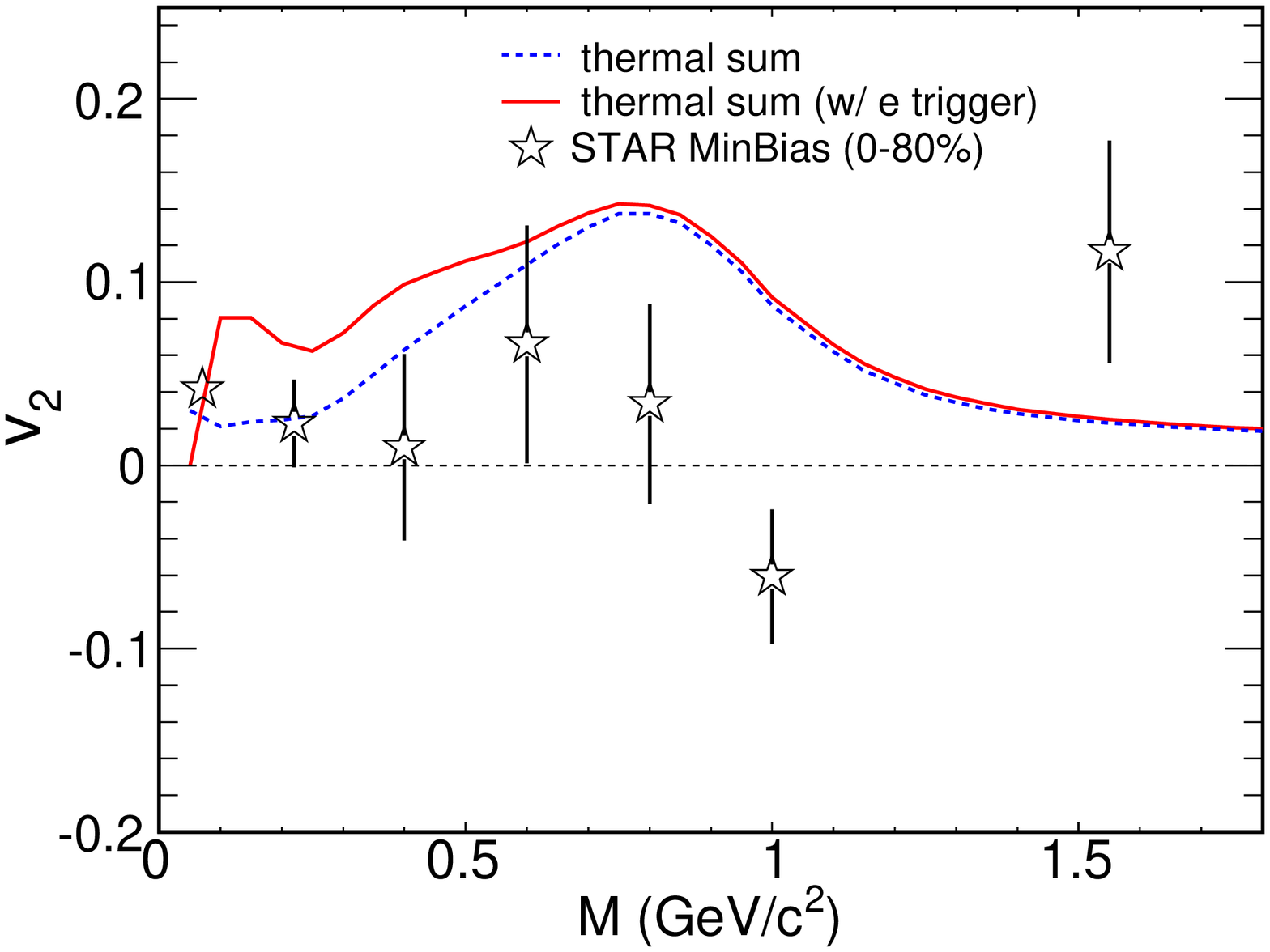}

\caption{\label{fig:v2MB} (Color Online) Thermal elliptic flow of dielectrons
from $\sqrt{s_{NN}}=200$~GeV Au+Au minimum bias collisions and comparison
with STAR data \cite{STAR_PRC}.}
\label{v2MB} 
\end{figure}

In Fig.\ref{fig:v2MB}, the red solid line is our calculated elliptic
flow of thermal di-electrons from Au+Au minimum bias collisions (0-80\%
centrality) at $\sqrt{s_{NN}}=200$~GeV. (The dashed line is the result
 before taking into account of the acceptence factor of single electron trigger.) 
As a reference, we also show STAR data \cite{STAR_PRC} (stars) of the elliptic flow of di-electrons
(from all sources, of course). The calculated elliptic flow of thermal
dielectrons is comparable, in fact even larger than data for all di-electrons.
The cocktail contribution will inherit a certain elliptic flow because
hadrons carry elliptic flow before decaying into di-electrons. Most
models predict it to be lower than the STAR reference \cite{Rapp_3,Vujanovic,Xu2},
whereas our result based on EPOS3 is larger.

To make a complete collection of our theoretical results, we present
the centrality dependence and the $n$ dependence of the flow harmonics
$v_{n}$ of dileptons in the following. 
\begin{figure*}
\includegraphics[scale=0.8]{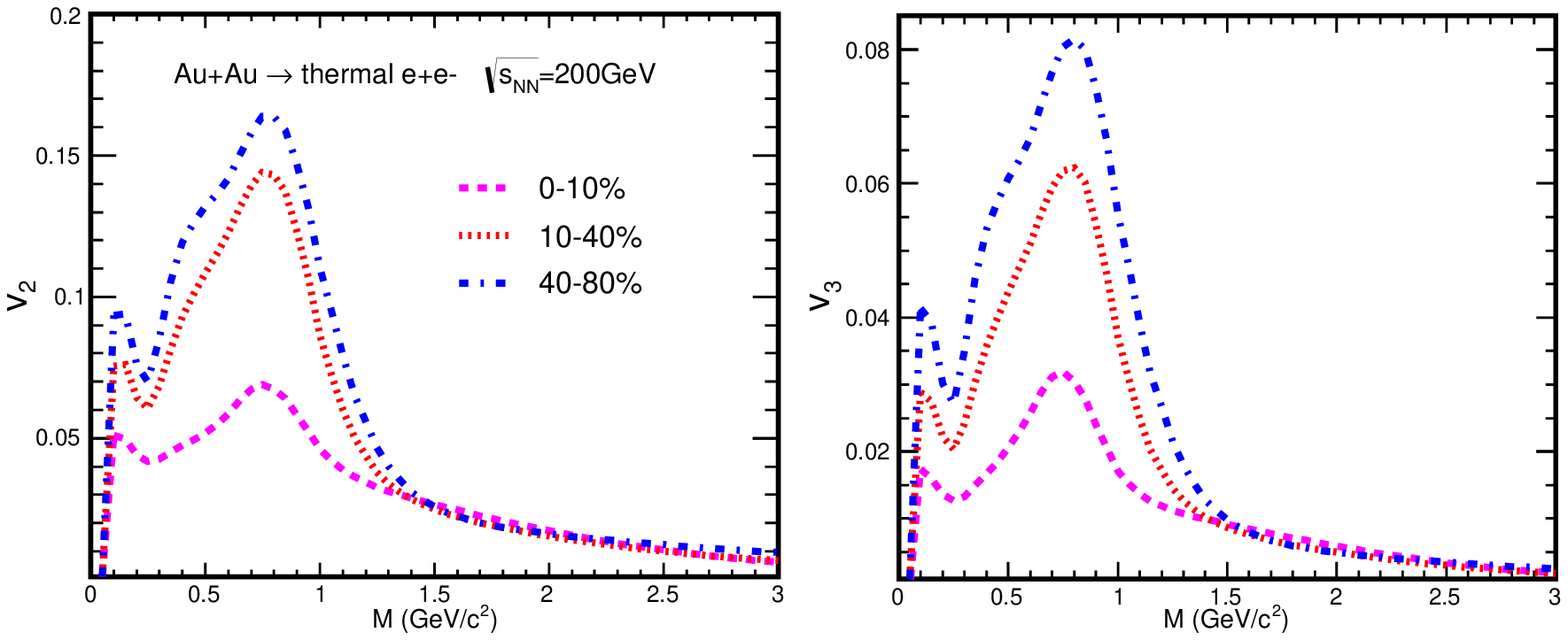}

\protect\caption{\label{fig:v2andv3} (Color Online) The elliptic flow (left) and trianger
flow (right) of thermal dielectrons from AuAu collisions at $\sqrt{s_{NN}}=200$~GeV.
0-10\%(pink dashed line), 10-40\%(red dotted line), 40-80\%(dark blue
dashed-dotted line).}
\end{figure*}

In Fig.\ref{fig:v2andv3} is shown the elliptic flow of thermal dielectrons
from AuAu collisions at $\sqrt{s_{NN}}=200$~GeV for different centralities:
0-10\%(pink dashed line), 10-40\%(red dotted line), 40-80\%(dark blue
dashed-dotted line). The centrality dependence of high order $n$
is also investigated. A strong centrality dependence, from central
to peripheral collisions, occurs not only to the elliptic flow but
also to higher order, $n=3,4,5$.

\begin{figure}
\includegraphics[scale=0.4]{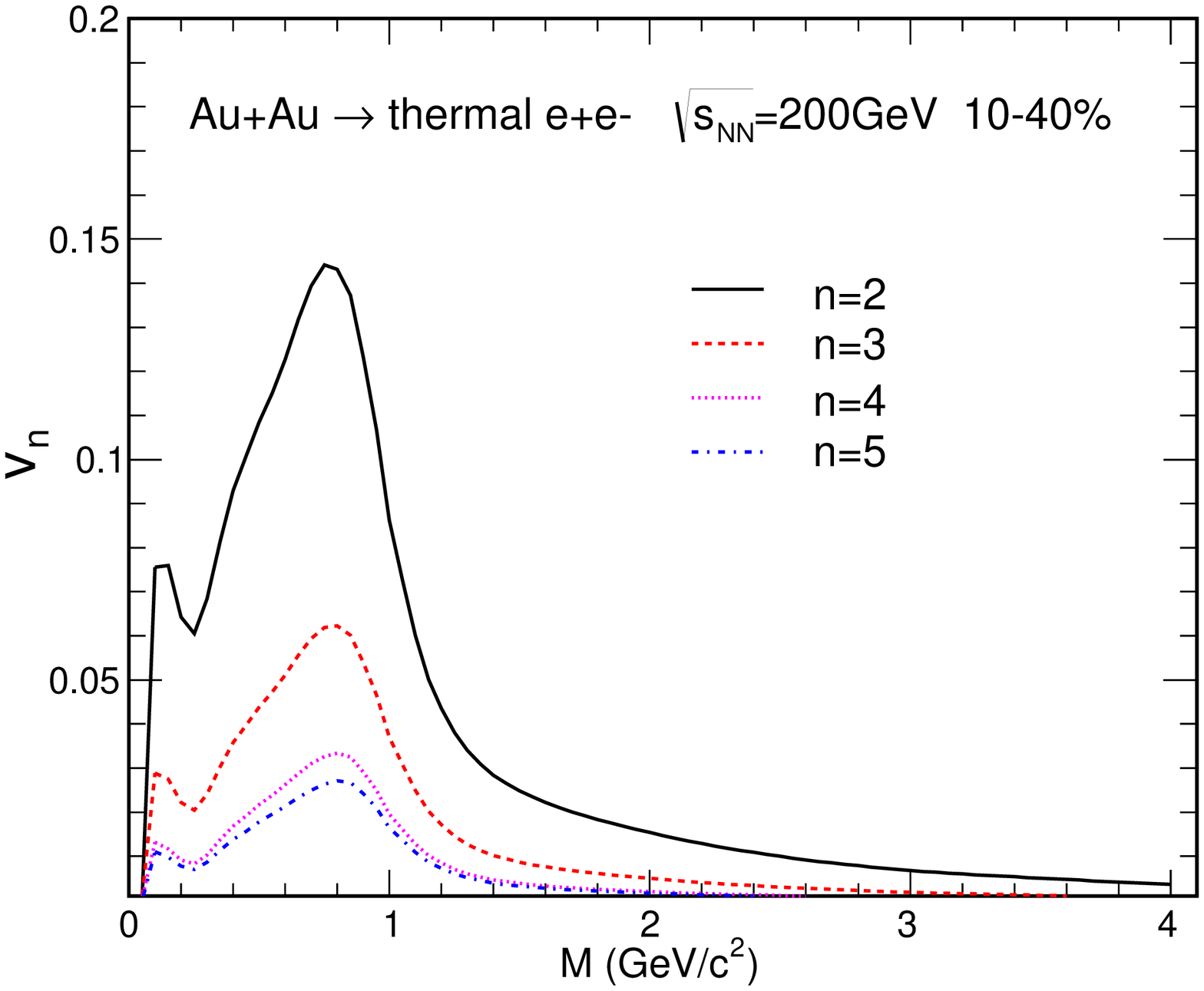}

\protect\caption{\label{fig:vn_10-40} (Color Online) Predicated harmonics coefficients
$v_{n}$ ($n=2,3,4,5$) of thermal dielectrons are shown by various
curves for centrality 10-40\%.}
\end{figure}

In Fig.\ref{fig:vn_10-40}, the higher order harmonics coefficients
$v_{n}$ ($n=2,3,4,5$) of thermal dielectrons from Au-Au collisions
at $\sqrt{s_{NN}}=200$~GeV with centrality of 10-40\% is presented.
The curves have same shape, same trend as direct photons. The magnitude
decreases monotonically with the order $n$. And it is the case for
all the investigated centrality class, 0-10\%, 10-40\% and 40-80\%.
Such a behavior was observed for charged hadrons at low $pt$ by ALICE~\cite{ALICE:2011ab}.

\section{DISCUSSION AND CONCLUSION}

We investigated the anisotropic emission of photons and dileptons from
Au+Au central collision at $\sqrt{s_{NN}}=200$~GeV at the RHIC based on a hadron data constrained 
(3+1) dimensional hydrodynamic model EPOS3. All parameters are the default values of EPOS3.102 for the 
hydrodynamic evolution of the model. 
Thermal photons and dileptons are assumed to emit from hydro initial time, with the full rates 
introduced in the approach section, but without viscous correction in either QGP phase or HG phase.

The anisotropic emission of charged hadrons from the model agrees to experimental data at different centrality classes. And a good reproduction of the measured direct photon elliptic flow is obtained for all centralities. Thus we predicted the elliptic flow of thermal dileptons, which is higher than the available results 
from other models, and comparable to the measured elliptic flow of total dileptons 
(thermal + cocktail) by STAR collaboration. 
We also made a prediction to high order flow coefficients of thermal dileptons, though our calculated 
triangular flow of direct photons does not coincide with PHENIX data so well. 
Since the triangular flow of charged hadrons calculated with cumulant approach agrees with data. and    
the deviation of triangular flow of direct photons from data has a centrality dependence, we should check 
whether the high order event planes of thermal emission defined in this way 
coincide with the experimental measurement in future. The event plane of second order defined so 
seems more reliable than those of higher order, due to the strong initial geometry.

The investigation of direct photons and dileptons from the same model, with a consistent description of 
the space-time evolution is useful. A underestimation of thermal photons from EPOS3.102 was found, 
compared with PHENIX measured transverse momentum spectra of direct photons. And a consistent 
underestimation of thermal dileptons was also found, compared with STAR measured invariant mass spectrum 
of dileptons. This shows a consistency between the two experimental measurements, and a requirement to 
improve the calculation for more thermal emissions.    
 
\begin{acknowledgments}
S. X. Liu thanks J. Zhao and H. J. Xu for very helpful discussion.
This work was supported by the Natural Science Foundation of China
under Project No.11275081 and by the Program for New Century Excellent
Talents in University (NCET). \end{acknowledgments}

\end{document}